\documentclass[aps,prb,floatfix,reprint,superscriptaddress]{revtex4-1}

\usepackage{graphicx}% Include figure files
\usepackage{dcolumn}% Align table columns on decimal point
\usepackage{bm}% bold math
\usepackage{amssymb}% bold math
\usepackage{rotating} %rotate figures and tables

\begin{document}

%\preprint{}
\title{Molecule-optimized Basis Sets and Hamiltonians for Accelerated Electronic Structure Calculations of Atoms and Molecules}

\author{Gergely Gidofalvi}

\affiliation{Department of Chemistry, Gonzaga University, Spokane, WA 99258}%

\author{David A. Mazziotti}

\email{damazz@uchicago.edu}

\affiliation{Department of Chemistry and The James Franck Institute, The University of Chicago, Chicago, IL 60637}%

\date{Submitted October 14, 2013; Revised December 8, 2013}

% It is always \today, today,
% but any date may be explicitly specified

\begin{abstract}

Molecule-optimized basis sets, based on approximate natural
orbitals, are developed for accelerating the convergence of quantum
calculations with strongly correlated (multi-referenced) electrons.
We use a low-cost approximate solution of the anti-Hermitian
contracted Schr{\"o}dinger equation (ACSE) for the one- and
two-electron reduced density matrices (RDMs) to generate an
approximate set of natural orbitals for strongly correlated quantum
systems.  The natural-orbital basis set is truncated to generate a
molecule-optimized basis set whose rank matches that of a standard
correlation-consistent basis set optimized for the atoms. We show
that basis-set truncation by approximate natural orbitals can be
viewed as a one-electron unitary transformation of the Hamiltonian
operator and suggest an extension of approximate natural-orbital
truncations through two-electron unitary transformations of the
Hamiltonian operator, such as those employed in the solution of the
ACSE.  The molecule-optimized basis set from the ACSE improves the
accuracy of the equivalent standard atom-optimized basis set at
little additional computational cost. We illustrate the method with
the potential energy curves of hydrogen fluoride and diatomic
nitrogen.   Relative to the hydrogen fluoride potential energy curve
from the ACSE in a polarized triple-zeta basis set, the ACSE curve
in a molecule-optimized basis set, equivalent in size to a polarized
double-zeta basis, has a nonparallelity error of 0.0154~a.u. which
is significantly better than the nonparallelity error of 0.0252~a.u.
from the polarized double-zeta basis set. \\

\indent {\bf Contact}: damazz@uchicago.edu (e-mail); 1-773-834-1762
(phone)

\end{abstract}

\maketitle

\section{Introduction}

While the basis sets describing atoms and molecules have been
extensively studied and optimized~\cite{Jensen,Szabo}, significant
opportunities exist for the improvement of molecule-optimized basis
sets and Hamiltonians for the acceleration of electronic structure
computations, especially in the presence of strong electron
correlation.   Standard atomic-orbital basis sets are optimized only
to minimize electronic energies of the constituent atoms rather than
the total electronic energy of the molecule~\cite{D89}.  While the
concept of atoms forming molecules has a critical role throughout
chemistry, such atom-centered basis sets do not capitalize on
opportunities for greater efficiency arising from the nature of the
bonding.  Molecule-optimized basis sets and Hamiltonians accelerate
correlation-energy calculations by minimizing the size (rank) of the
orbital basis set in the description of the correlated Hamiltonian
and wave function (or reduced density matrix). Such basis sets
realize the intuitive idea that the optimal basis set for a molecule
at a stretched geometry is different from the optimal basis set at
the equilibrium geometry.  Molecule-optimized orbitals, largely
based on approximate natural orbitals~\cite{RDM07,CB00,L55,M60,
C63,SL59,BD66,CSM79,W60,K64,DJ62,D72,B68,HL59,S59,RD66,RD69,HS63,LS69,
K69,S69,BLS65,NH63,TRPB77,BS67,FS03,MZP93,LS85,SH74,L64,L69,ALS75,
H73,M73,H73,MYN96,BP89,LK73} or frozen natural orbitals~\cite{BD70,
EK66,SGT89,AS04,TB05,KO06,TB08,LKD10,HS10,NHL09,SKS11,RK11,GBM11,DS13,
KSX13,AB87,PNK06,NPU05,A10,PAH11,KPH12,AS03,BR09,ATG09,LM12,CP12,TB05,
LKD10,HS10,DS13}, have been extensively studied in the context of
perturbation methods about the Hartree-Fock reference wave function,
but their study has been much more limited for strongly correlated
molecular systems.

The aim of the present paper is to generate a molecule-optimized
double-zeta basis set and Hamiltonian in which quantum chemistry
calculations of strongly correlated systems can be performed.  We
(i) form the molecule-optimized double-zeta basis set and
Hamiltonian from an {\em approximate} solution of the anti-Hermitian
contracted Schr{\"o}dinger equation (ACSE)~\cite{AC06,
*AC07B,*AC07C,*AC07D,ACMR,*GM09b,*RFM09,ACV07,*ACV09,AC08,FRM09,GM10,
SM10,*SM11,*G11,*SM11,*SM12} in a higher (triplet-zeta) basis set
and (ii) apply the molecule-optimized double-zeta basis set and
Hamiltonian to solving the ACSE.  The molecule-optimized double-zeta
orbitals capture strong correlation, if present, because they are
generated from an ACSE calculation starting with an initial
multi-configuration self-consistent-field (MCSCF) two-electron
reduced density matrix (2-RDM).  In the theory section we present a
general formulation for molecule-optimized basis sets and
Hamiltonians in terms of unitary transformations of the Hamiltonian
operator.  Optimized orbitals can be viewed as one-electron unitary
transformations of the Hamiltonian operator.  Within this framework
more general unitary transformations can potentially be explored.
Illustrative applications are made to the potential energy curves of
hydrogen fluoride and diatomic nitrogen.  The ACSE-computed
molecule-optimized basis sets significantly improve the
non-parallelity errors in both curves.  Although we specifically use
the optimized basis orbitals and Hamiltonians in the ACSE, they can
be more generally employed with any electronic structure method.

\section{Theory}

Molecule-optimized orbitals are generally developed from approximate
natural orbitals in section~\ref{sec:no}.  The construction of a set
of natural orbitals from the ACSE that are suitable for treating
strongly correlated many-electron molecular systems is described in
section~\ref{sec:acse}.   In section~\ref{sec:ham} we recast the
acceleration of convergence as a unitary transformation of the
molecular Hamiltonian and suggest extensions of the approximate
natural-orbital transformations.

\subsection{Approximate natural orbitals}

\label{sec:no}

The ``best'' molecule-optimized molecular orbitals are the natural
orbitals, the eigenfunctions of the 1-RDM.  The optimality of the
natural orbitals follows from a mathematical theorem derived by E.
Schmidt in 1907~\cite{C63,CB00,E07}.  While finding the exact
natural orbitals in a large standard atom-optimized basis set might
require the same computational cost as solving the correlation
problem in that large basis set, significant cost savings can be
achieved by identifying an approximate set of natural orbitals and
then solving the correlation problem in a truncated set of these
orbitals.

Approximate natural orbitals can be obtained from a low-cost
correlation-energy calculation and then employed after truncation in
a higher cost correlation-energy calculation.  Examples of the
strategy from the literature include the early use of natural
orbitals from perturbation theory~\cite{L64,BLS65,EK66,H73,M73,SH74}
or iterative refinement~\cite{D72,TRPB77} in configuration
interaction and the recent use of natural orbitals from second-order
many-body perturbation theory in coupled cluster
calculations~\cite{TB05,LKD10,HS10,DS13}.  Most previous
calculations differ from the general approach to the optimal natural
orbitals adopted here in two respects: (1) they typically employ a
truncation  scheme for the natural orbitals based on a threshold for
their  occupations and (2) they usually determine approximate
natural  orbitals either from or for single-reference electron
correlation  methods.  Taube and Bartlett~\cite{TB05,TB08} have
truncated their natural orbitals according to a predefined
percentage, and Roos and co-workers~\cite{ATG09} have employed
approximate natural orbitals in complete-active-space second-order
perturbation theory.

In this work we generate molecule-optimized basis sets that use a
truncation of the natural orbitals based on the rank of the orbitals
(see also Ref.~\cite{AS03} for a truncation by basis-set size).  For
example, the approximate natural orbitals are obtained from a
low-cost method in a large standard atom-optimized basis set
\begin{equation}
\label{eq:D1eig} ^{1} D  v_{i} = n_{i} v_{i},
\end{equation}
where ${}^{1} D$ denotes the 1-RDM, $n_{i}$ are the natural
occupation numbers ordered from largest to smallest, and $v_{i}$ are
the eigenvectors whose components denote the expansion coefficients
of the natural orbitals in terms of the initial molecular-orbital
basis set. Then the set of natural orbitals $\{ v_{i} \}$ is
truncated to share the rank $M$ of the smaller standard
atom-optimized basis set. In accordance with the Schmidt theorem,
the largest $M$ of the $n_{i}$ are retained to generate the optimal
set of $M$ orbitals. The compact molecule-optimized basis set can
then be employed in a higher cost method for more accurate and more
efficient description of the molecule's electron correlation.

Truncation by basis-set rank has a different philosophy from
truncation by threshold. In truncation by threshold the aim of the
calculation is to reproduce the accuracy of the larger basis set
within a given tolerance (threshold), but in truncation by basis-set
rank the aim of the calculation is to attain some of the accuracy of
the larger basis set at the significantly reduced cost of a smaller
basis set.  For larger molecules where the computational cost of the
larger basis set is prohibitive, the strategy of truncation by
basis-set rank has important advantages because {\em the basis-set
rank can be chosen to remain within existing computational
resources}. Furthermore, the generation of molecule-optimized basis
sets which mimic traditional atom-optimized basis sets share some
advantages of atom-optimized basis sets such as correlation
consistency and systematic extrapolation to the complete-basis-set
limit.

Secondly, as discussed in section~\ref{sec:acse}, we aim to develop
molecule-optimized molecular orbitals that can be employed in
multi-reference calculations for the description of strongly
correlated electrons.   We generate approximate natural orbitals
through a partial solution of the ACSE, starting with a 2-RDM from
an MCSCF calculation.  The resulting natural orbitals from the
partial ACSE solution have a natural ordering with respect to
correlation effects that inherently require multiple many-electron
configurations in the reference wave function.   Natural orbitals
from single-reference theories typically do not reflect the
multi-reference correlation in the wave functions of highly
correlated atoms molecules.  Furthermore, as shown in the results,
canonical orbitals from MCSCF, ordered by their canonical energies,
do not provide a suitable ordering for accelerating convergence with
respect to basis-set size.

\subsection{ACSE natural orbitals}

\label{sec:acse}

Solution of the anti-Hermitian contracted Schr{\"o}dinger equation
(ACSE)~\cite{AC06,*AC07B,*AC07C,*AC07D,ACMR,*GM09b,*RFM09,ACV07,*ACV09},
the anti-Hermitian part of the contracted Schr{\"o}dinger equation
(CSE)~\cite{V93,*V93b,NY96,*NY97,M98a,*M98b}, for the 2-RDM and its
energy can be tuned for single-reference or multi-reference electron
correlation through the choice of the initial 2-RDM.  The 2-RDM can
be chosen from an initial mean-field (Hartree-Fock) or a correlated
calculation such as a multi-configuration self consistent field
(MCSCF) calculation~\cite{AC06,*AC07B,*AC07C,*AC07D,ACMR,
*GM09b,*RFM09}. The ACSE method is applicable to both ground and excited
states as well as arbitrary spin states~\cite{ACMR,*GM09b,*RFM09}.
It has been applied to studying multi-reference correlation in
excited states and conical intersections in the photoexcitation of
{\em gauche}-1,3-butadiene to form bicyclobutane~\cite{SM11}, the
tautomerization of vinyl alcohol to acetylaldehyde~\cite{SM12}, and
the reaction of firefly luciferin for bioluminescence~\cite{G11}.

In a finite basis set the contracted Schr{\"d}inger equation (CSE)
as well as its anti-Hermitian part (ACSE) can expressed in second
quantization as
\begin{equation}
\label{eq:CSE2} \langle \Psi_{n} | {\hat a}^{\dagger}_{i} {\hat
a}^{\dagger}_{j} {\hat a}_{l} {\hat a}_{k} {\hat H} | \Psi_{n}
\rangle = E_{n} \hspace{1mm} {}^{2} D^{i,j}_{k,l}
\end{equation}
and
\begin{eqnarray}
\label{eq:ACSE2} \frac{1}{2} \langle \Psi_{n} |  [ {\hat
a}^{\dagger}_{i} {\hat a}^{\dagger}_{j} {\hat a}_{l} {\hat a}_{k},
{\hat H} ] | \Psi_{n} \rangle & = & 0 ,
\end{eqnarray}
where each index $i$, $j$, $k$, and $l$ denotes a one-electron spin
orbital that is a product of a spatial orbital and a spin function
$\sigma$ equal to either $\alpha$ (+1/2) or $\beta$ (-1/2) and the
elements of the 2-RDM
\begin{equation}
\label{eq:2rdm} {}^{2} D^{i,j}_{k,l} = \langle \Psi_{n} | {\hat
a}^{\dagger}_{i} {\hat a}^{\dagger}_{j} {\hat a}_{l} {\hat a}_{k} |
\Psi_{n} \rangle
\end{equation}
follow from the expectation value of the 2-RDO with respect to $|
\Psi_{n} \rangle$.  In second quantization the creation operator
$a^{\dagger}_{i}$ generates an electron in the $i^{\rm th}$ spin
orbital while the annihilation operator $a_{k}$ destroys an electron
in the $k^{\rm th}$ spin orbital.  For a quantum many-electron
system the Hamiltonian is expressible as
\begin{equation}
\label{eq:H} {\hat H} = \sum_{p,s}{ {}^{1} K^{p}_{s} {\hat
a}^{\dagger}_{p} {\hat a}_{s} } + \sum_{p,q,s,t}{{}^{2}
V^{p,q}_{s,t} {\hat a}^{\dagger}_{p} {\hat a}^{\dagger}_{q} {\hat
a}_{t} {\hat a}_{s} }
\end{equation}
where the one- and two-electron reduced Hamiltonian matrices ${}^{1}
K$ and ${}^{2} V$ contain the one- and two-electron integrals
respectively.  By rearranging the creation and annihilation
operators according to the anti-commutation relations for fermions,
we can write the CSE in terms of the elements of the 2-, 3-, and
4-RDMs and the ACSE in terms of the elements of the 2- and 3-RDMs.
Explicit expressions for these contracted equations in terms of the
spin-orbital elements of the reduced Hamiltonians and RDMs are given
elsewhere~\cite{AC06,*AC07B,*AC07C,*AC07D,ACMR,*GM09b,*RFM09,ACV07,*ACV09}.

The ACSE can be solved by propagating the following initial-valued
differential equation as a function of the parameter $\lambda$ which
serves as an imaginary time:
\begin{equation}
\label{eq:dD} \frac{d \hspace{1mm} {}^{2} D^{i,j}_{k,l}}{d \lambda}
= \langle \Psi(\lambda) | [ {\hat a}^{\dagger}_{i} {\hat
a}^{\dagger}_{j} {\hat a}_{l} {\hat a}_{k}, {\hat S}(\lambda) ] |
\Psi(\lambda) \rangle
\end{equation}
where the two-body operator ${\hat S}$
\begin{equation}
\label{eq:S} \hat{S}(\lambda) = \sum_{p,q,s,t} {}^{2} S^{p,q}_{s,t}
(\lambda) {\hat a}^{\dagger}_{p} {\hat a}^{\dagger}_{q} {\hat a}_{t}
{\hat a}_{s}
\end{equation}
depends upon a two particle reduced matrix $^{2} S$ equal to the
residual of the ACSE
\begin{equation}
\label{eq:S2} {}^{2} S^{p,q}_{s,t}(\lambda) = \langle \Psi(\lambda)
| [ {\hat a}^{\dagger}_{p} {\hat a}^{\dagger}_{q} {\hat a}_{t} {\hat
a}_{s}, {\hat H} ] | \Psi(\lambda) \rangle
\end{equation}
where ${\hat H}$ is the Hamiltonian operator.  The dependence of the
above equations on the three-electron reduced density matrix (3-RDM)
is removed by reconstructing the 3-RDM as a cumulant functional of
the lower 1-and 2-RDMs~\cite{M98b,*M98c,*M99a,*KM99}.   The 2-RDM is
propagated until either the energy or the residual the ACSE ceases
to decrease.

Because the ACSE can treat multi-reference correlation, it can serve
as a general platform for creating an approximate set of natural
orbitals. The 1-RDM is obtainable from the 2-RDM by contraction
\begin{equation}
^{1} D^{i}_{k} = \frac{1}{N-1} \sum_{j}{ {}^{2} D^{i,j}_{k,j} },
\end{equation}
and the natural orbitals and their occupations are readily obtained
from Eq.~(\ref{eq:D1eig}).  A family of approximate orbitals can be
systematically generated from the ACSE by evolving the 2-RDM over a
short length in the parameter $\lambda$.  By choosing the distance
in $\lambda$ to be a small fraction of the total distance
$\lambda^{*}$ required for the solution of the ACSE, we can obtain
an approximate set of natural orbitals at low computational cost.
The evolution over the short distance in $\lambda$ can be performed
in a large standard atom-centered basis set.  From the 1-RDM
obtained, a truncated set of natural orbitals sharing the rank $M$
of the smaller standard atom-centered basis set can be employed for
solving the ACSE to convergence.  Hence, through the choice of the
evolution distance in $\lambda$ we are able to generate both
low-cost and higher cost methods for electron correlation directly
within a common ACSE framework.

For convenience, we diagonalize only the virtual-virtual block of
the 1-RDM to obtain natural orbitals in terms of the virtual MCSCF
orbitals; in this manner, we can truncate these approximate natural
orbitals without changing the original MCSCF 1-RDM at $\lambda=0$.
These approximate natural orbitals are similar in spirit to those
created from the Hartree-Fock virtual-virtual block of the 1-RDM in
single-reference methods, which have been called frozen natural
orbitals~\cite{EK66, BD70,SGT89,DS13}. Importantly, because the
approximate natural orbitals obtained from the ACSE are molecule
optimized, they incorporate important features of the molecule's
electron density and chemical bonding that are not present in the
standard atom-centered basis sets of the same size (or rank).

\subsection{Molecule-optimized Hamiltonians}

\label{sec:ham}

The generation of molecule-optimized basis sets through the use of
natural orbitals and Schmidt's theorem can also be viewed as a
unitary transformation of the Hamiltonian in the original larger
basis set ${\hat H}_{0}$ to produce a more compact Hamiltonian
${\hat H}_{1}$ whose non-negligible elements can be captured in a
smaller basis set
\begin{equation}
{\hat H}_{1}= e^{-{\hat S}_{1}} {\hat H_{0}} e^{{\hat S}_{1}},
\end{equation}
where ${\hat S}_{1}$ is a one-body anti-Hermitian operator
\begin{equation}
\label{eq:S1} \hat{S}_{1} = \sum_{p,s} {}^{1} S^{p}_{s} {\hat
a}^{\dagger}_{p} {\hat a}_{s} .
\end{equation}
Similar anti-Hermitian operators arise in the unitary
transformations underlying contracted Schr{\"o}dinger
theory~\cite{RDM07,V93,*V93b,NY96, *NY97, M98a,*M98b} including the
solution of the ACSE~\cite{AC06, *AC07B,*AC07C,*AC07D,
ACMR,*GM09b,*RFM09,ACV07,*ACV09,AC08,FRM09, GM10,SM10,*SM11,
*G11,*SM11,*SM12}.  In the ACSE we employ unitary transformations
from not only one-body anti-Hermitian operators but also such
transformations from two-body anti-Hermitian operators, which are
critical to capturing important many-body correlation effects.

The molecule-optimized Hamiltonian from a one-body unitary
transformation can be generalized to a molecule-optimized
Hamiltonian from a two-body unitary transformation
\begin{equation}
{\hat H}_{2}= e^{-{\hat S}_{2}} {\hat H_{0}} e^{{\hat S}_{2}},
\end{equation}
where ${\hat S}_{2}$ is a two-body anti-Hermitian operator.  As in
the previous case, the two-body transformation produces a more
compact Hamiltonian ${\hat H}_{2}$ whose non-negligible elements can
be captured in a smaller basis set.  Because the two-body unitary
transformations contain the one-body unitary transformations, the
set of potential Hamiltonian operators $\{{\hat H}_{2}\}$  is larger
than a set of potential Hamiltonian operators $\{{\hat H}_{1}\}$ .
Consequently, the two-body transformations generalize the set of
molecule-optimized Hamiltonians obtainable from approximate natural
orbitals.

Unlike the one-body transformations, the two-body transformations
generate three-body Hamiltonians whose expectation values depend
upon the three-electron RDM (3-RDM).  As in contracted
Schr{\"o}dinger theory, however, these three-body Hamiltonians can
be readily approximated as two-body Hamiltonians through cumulant
reconstruction of the three-electron reduced density
operators~\cite{M98b, *M98c,*M99a,*KM99}.  While these extensions of
the natural-orbital transformations are not pursued in the present
work, the ACSE theory~\cite{AC06, *AC07B,*AC07C,*AC07D,ACMR,*GM09b}
in the previous section provides a useful framework for (i)
approximating suitable ${\hat S}_{2}$ operators  and (ii) recasting
the Hamiltonians ${\hat H}_{2}$ as two-body operators through
cumulant reconstruction.  Recently, related two-body transformations
of the Hamiltonian with cumulant reconstruction have been employed
in the context of an explicit $r12$ theory~\cite{YS12}.

\section{Applications}

After brief discussion of the computational methodology, we apply
the molecule-optimized basis sets described in the previous section
to generating potential energy curves from hydrogen fluoride and
diatomic nitrogen.

\subsection{Computational methodology}

The initial MCSCF 2-RDM is computed with the wave function from an
MCSCF calculation in the GAMESS package for electronic
structure~\cite{GAMESS}.  The ACSE calculations are performed with
the code developed by one of the authors in Refs.~\cite{AC06,
*AC07B,*AC07C,*AC07D,ACMR,*GM09b,*RFM09}.  Approximate sets of natural
orbitals are generated from ACSE calculations in the
correlation-consistent polarized valence triple-zeta (TZ) basis
sets~\cite{D89}.  Unless stated otherwise, the 2-RDM is evolved from
$\lambda = 0.0$ to $\lambda = 0.01$.  This evolution is a small
fraction of the total evolution in $\lambda$ from $0.0$ to
$\lambda^{*}$ required for satisfying the ACSE method's convergence
criteria.  As shown in previous work~\cite{AC06}, convergence
typically occurs at a value $\lambda^{*}$ between 1 and 10. The
resulting natural orbitals are then truncated based on orbital
occupations to produce a molecule-optimized basis set whose rank $M$
equals that of the standard correlation-consistent polarized valence
double-zeta (DZ) basis set~\cite{D89}. The ACSE is then evolved in
this molecule-optimized basis set until convergence.

\subsection{Hydrogen fluoride}

\begin{figure}[htp!]

\includegraphics[scale=0.4]{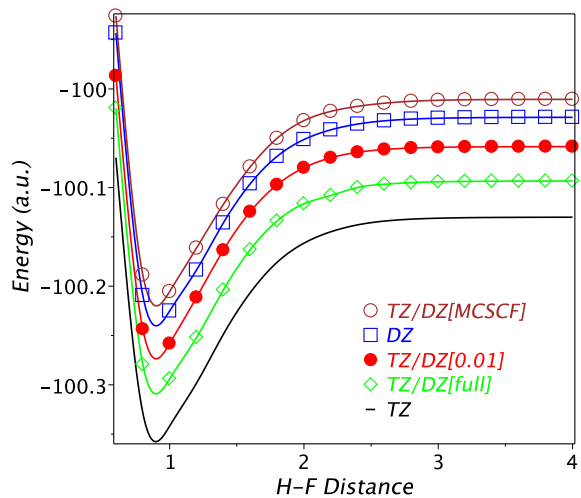}

\caption{The potential energy curve in the molecule-optimized basis
set from the ACSE with $\lambda$ equal to 0.01 (TZ/DZ[0.01]) is
compared to those from solving the ACSE in the standard correlation
consistent basis sets, DZ and TZ, the nonstandard DZ basis set
derived from the energy-ordered orbitals of MCSCF in the TZ basis
set (TZ/DZ[MCSCF]) as well as the molecule-optimized basis set with
$\lambda$ evolved its full distance $\lambda^{*}$ to convergence
(TZ/DZ[full]). Relative to TZ, the nonparallelity error (NPE) of
0.0154~a.u. from TZ/DZ[0.01] is significantly better than the error
of 0.0252~a.u. from DZ or the error of 0.02411~a.u. from
TZ/DZ[MCSCF].}

\label{f:hf}

\end{figure}

\begin{table*}[ht!]

\caption{TABLE 1: Relative to TZ, the table reports the energy
errors from the ACSE from the standard atom-optimized basis set DZ
as well as a series of molecule-optimized basis sets for $\lambda$
equal to 0.01, 0.05, 0.10, and $\lambda^{*}$ where $\lambda^{*}$
represents the full $\lambda$ trajectory to convergence.  The
results show the error relative to TZ continues to decrease as the
approximate set of natural orbitals are improved through longer
$\lambda$ evolutions. The molecule-optimized basis set from
$\lambda$ equal to 0.01 (TZ/DZ[0.01]) offers an improvement in
accuracy at a computational cost that is not significantly different
from that of the standard DZ calculation.}

\label{t:t}

\begin{ruledtabular}
\begin{tabular}{ccccccc}

Bond & Energy (a.u.) & \multicolumn{5}{c}{Energy Errors (a.u.)} \\
\cline{2-2} \cline{3-7}

Distance (\AA) & TZ & DZ & TZ/DZ[0.01] & TZ/DZ[0.05] & TZ/DZ[0.10] & TZ/DZ[full] \\
\hline

0.8 & -100.328456 & 0.119026 & 0.084854 & 0.077534 & 0.065513 & 0.049063 \\

1.0 & -100.341785 & 0.116656 & 0.083603 & 0.077410 & 0.063675 & 0.047857 \\

1.2 & -100.298315 & 0.114687 & 0.086913 & 0.075308 & 0.060738 & 0.046282 \\

1.4 & -100.248797 & 0.112944 & 0.085235 & 0.079852 & 0.059025 & 0.044956 \\

1.8 & -100.175789 & 0.107404 & 0.078613 & 0.072371 & 0.061938 & 0.041952 \\

2.2 & -100.144837 & 0.103344 & 0.075038 & 0.069548 & 0.058482 & 0.036774 \\

2.8 & -100.132043 & 0.101452 & 0.072117 & 0.066604 & 0.055794 & 0.037054 \\

3.4 & -100.130323 & 0.101302 & 0.071648 & 0.066151 & 0.055370 & 0.036750 \\

\end{tabular}
\end{ruledtabular}

\end{table*}

The hydrogen fluoride molecule with its highly polarized chemical
bond has contributions from multiple configurations in the
dissociative region of its potential energy curve.  Here we generate
the potential energy curve in the molecule-optimized basis set from
the ACSE with $\lambda$ equal to 0.01. In Fig.~1 this potential
energy curve is compared to those from solving the ACSE in the
standard correlation-consistent basis sets, DZ and TZ, the
nonstandard DZ basis set derived from the energy-ordered orbitals of
MCSCF in the TZ basis set (TZ/DZ[MCSCF]), as well as the
molecule-optimized basis set with $\lambda$ evolved its full
distance $\lambda^{*}$ to convergence (TZ/DZ[full]). Importantly,
even though the molecule-optimized basis set with $\lambda=0.01$ has
a rank equal to that of the polarized basis set DZ, it improves the
energies from DZ by 20\% relative to the TZ basis set. Furthermore,
it has a nonparallelity error (NPE) of 0.0154~a.u. relative to TZ
which is significantly better than the error of 0.0252~a.u. from DZ
or the error of 0.02411~a.u. from TZ/DZ[MCSCF]. The NPE is defined
as the difference between the maximum and minimum errors in the
potential energy curve.  The NPE of the molecule-optimized basis set
with $\lambda=\infty$ at 0.0136~a.u. is not much different from that
of the basis set with $\lambda=0.01$.

Table~1 reports the energy errors from the ACSE relative to TZ from
the standard atom-optimized basis set DZ as well as a series of
molecule-optimized basis sets for $\lambda$ equal to 0.01, 0.05,
0.10, and $\lambda^{*}$ where $\lambda^{*}$ represents the full
$\lambda$ trajectory to convergence.  The results show the error
relative to TZ continues to decrease as the approximate set of
natural orbitals is improved through longer $\lambda$ evolutions.
Qualitatively, the space spanned by the $M$ most occupied natural
orbitals improves the energy with increasing $\lambda$ because it
better represents the part of the one-electron Hilbert space that
describes the electron density of the molecule.  This increasing
accuracy, however, comes at the price of increasing computational
cost.  The molecule-optimized basis set from $\lambda$ equal to 0.01
(TZ/DZ[0.01]) offers an improvement in accuracy at a computational
cost that is {\em not significantly different from that of the
standard DZ calculation}. In this case, the TZ/DZ[0.01] calculation
is {\em more than an order of magnitude faster} than the TZ
calculation.

\subsection{Diatomic nitrogen}

\begin{figure}[htp!]

\includegraphics[scale=0.4]{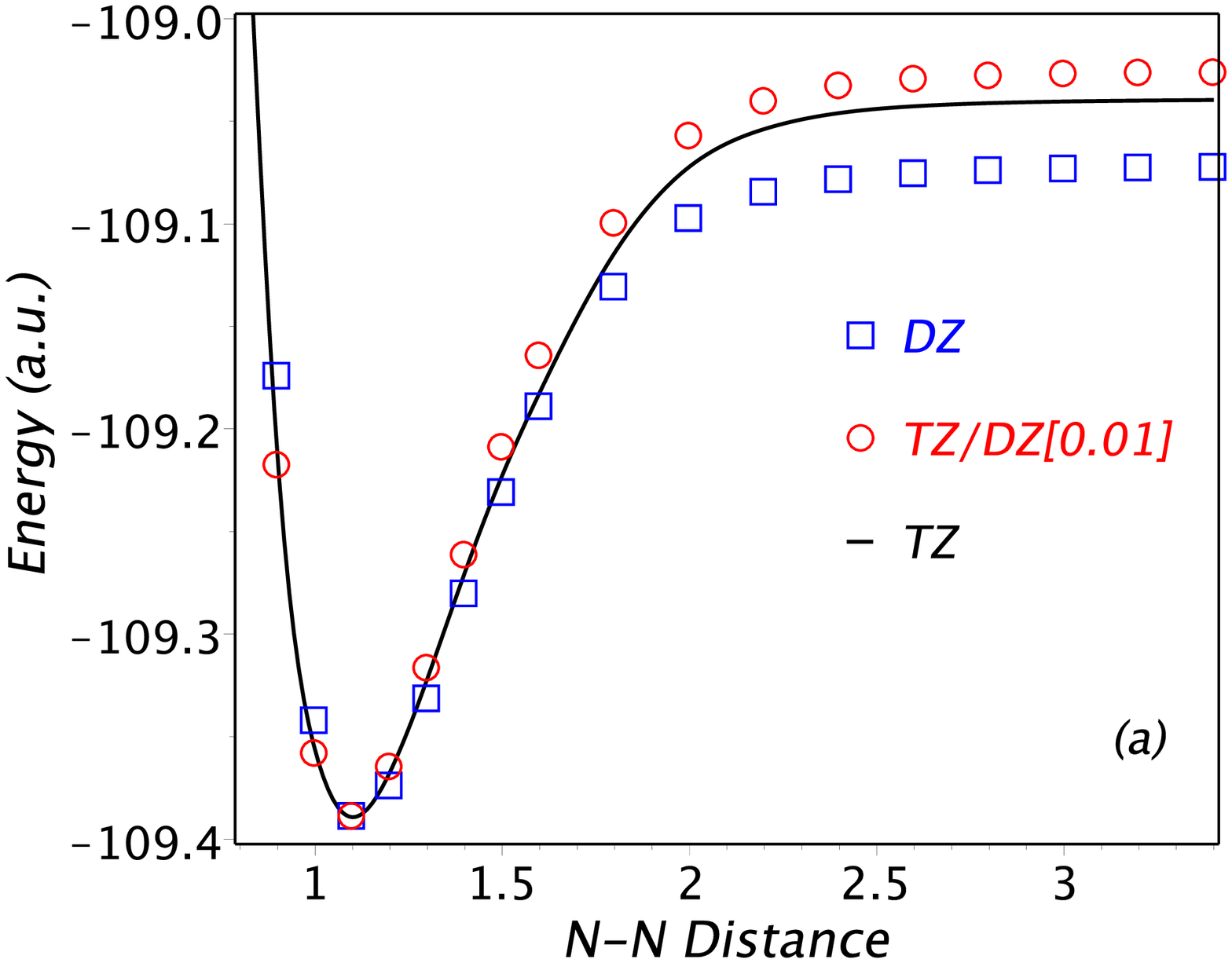}

\includegraphics[scale=0.4]{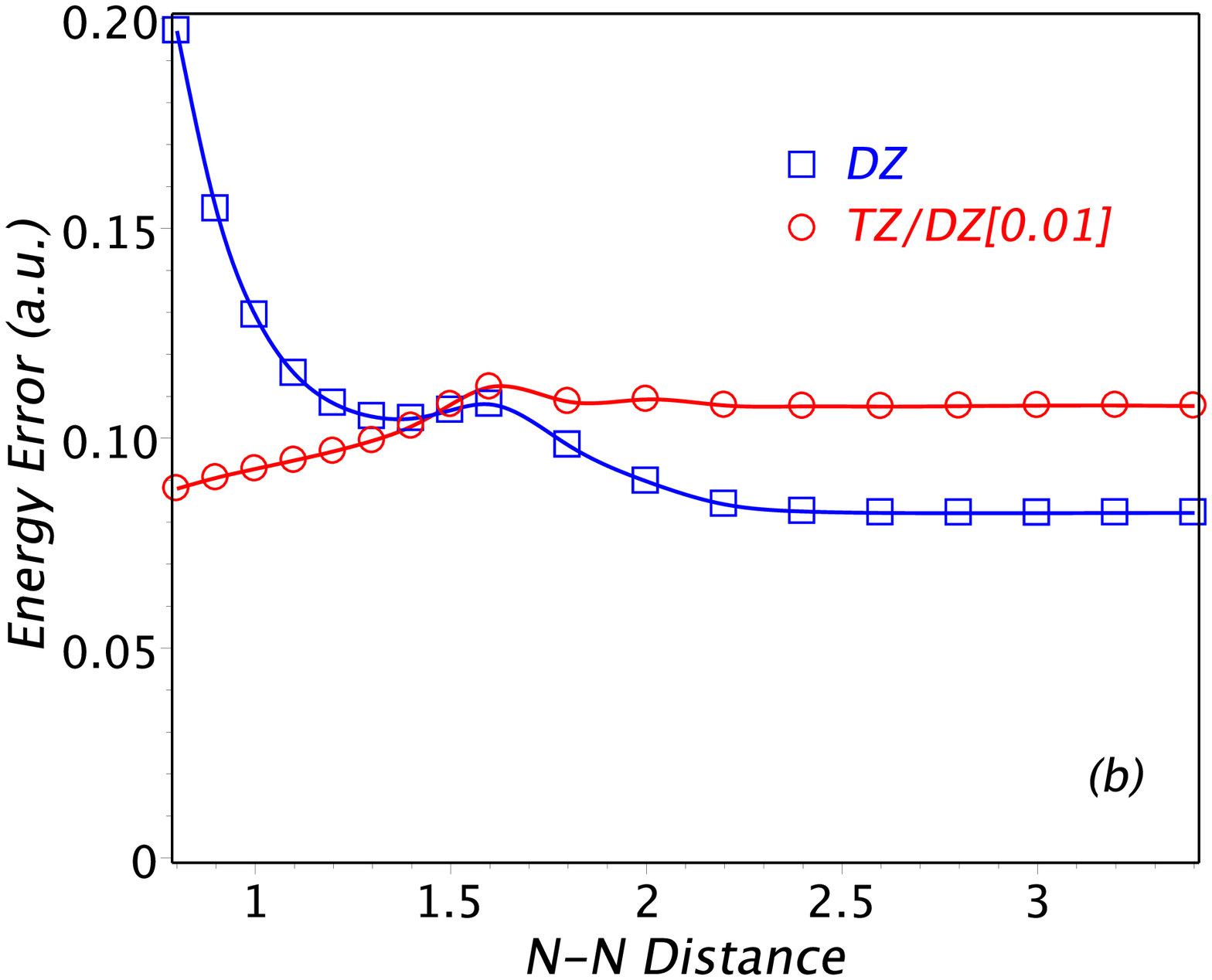}

\caption{For the dissociation of diatomic nitrogen with the ACSE the
figure compares the shape of the potential energy curve in the
molecule-optimized basis set with $\lambda$ equal to 0.01
(TZ/DZ[0.01]) to the shapes of potential curves from the standard DZ
and TZ basis sets.  The curves DZ and TZ/DZ[0.01] in part (a) are
shifted by $-0.115426$ and $-0.094708$~a.u. respectively to agree
with the energy from TZ at $1.1$~\AA.  Relative to TZ, the
TZ/DZ[0.01] curve better approximates both the curvature about
equilibrium and the dissociation energy than DZ; it significantly
improves the nonparallelity error of 0.115~a.u. of DZ to 0.024~a.u.
Part (b) shows the energy errors from DZ and TZ/DZ[0.01] relative to
TZ.}

\label{f:n2}

\end{figure}

Breaking the triple bond of diatomic nitrogen provides a challenging
problem for single-reference methods and a benchmark problem for
multi-reference methods. Here we generate the potential energy curve
for the nitrogen dissociation from the ACSE in the
molecule-optimized basis set with $\lambda$ equal to 0.01. The shape
of this potential energy curve is compared to the shapes of those
from the standard DZ and TZ basis sets in Fig.~2a.  The curves DZ
and TZ/DZ[0.01] in Fig.~2a are shifted by $-0.115426$ and
$-0.094708$~a.u. respectively to agree with the energy from TZ at
$1.1$~\AA.  Even though the molecule-optimized basis set has the
same computational cost as the DZ basis set, it significantly
improves the nonparallelity error relative to TZ from 0.115~a.u.
(DZ) to 0.024~a.u.  The curve from the molecule-optimized basis set
(TZ/DZ[0.01]) better approximates both the curvature about
equilibrium and the dissociation energy relative to TZ.

Figure~2b shows the energy errors from DZ and the molecule-optimized
basis set TZ/DZ[0.01] relative to TZ.  In terms of absolute
energies, TZ/DZ[0.01] improves the energies from DZ at bond lengths
in the vicinity of the equilibrium geometry; however, for bond
lengths greater than 1.6~\AA\ the molecule-optimized basis set
yields energies that are higher than those from the standard
correlation-consistent DZ basis set. This result can be understood
from recalling that the standard basis sets are optimized to
minimize atomic energies in the configuration interaction
singles-doubles method.  Upon dissociation the nitrogen molecule
breaks up into two nitrogen atoms, and hence, the standard basis set
is highly optimized in this region of the potential energy surface.
Examining the errors in the cc-pVDZ basis set relative to the
cc-pVTZ basis set, however, reveals that the errors at short bond
lengths are significantly larger than the errors at longer bond
lengths. This discrepancy in accuracy contributes to a large
nonparallelity error. The molecule-optimized basis set significantly
decreases this error by improving the energies in the equilibrium
region while sacrificing the accuracy of energies in the
dissociation region.  The basis set that is optimized for the
molecule provides a more balanced description of the molecule's
electron correlation throughout the potential energy surface.

\section{Discussion and conclusions}

Molecule-optimized basis sets have been presented for accelerating
the convergence of electron correlation calculations.  As in
previous work, the definition of the molecule-optimized basis set
depends upon the generation of an approximate set of natural
orbitals. Significant computational acceleration can be achieved
because the natural orbitals provide the optimal one-electron basis
set for the convergence of the many-electron wave function (or
two-electron reduced density matrix)~\cite{L55,C63}.  In contrast to
most previous work~\cite{SGT89,AS04,TB05,KO06,TB08,LKD10,HS10,NHL09,
SKS11,RK11,GBM11,DS13,KSX13,AB87,PNK06,NPU05,A10,PAH11,KPH12}, the
molecule-optimized basis sets (1) are defined by truncation of the
natural orbitals to a fixed rank that equals the rank of a standard
correlation-consistent polarized basis set and (2) are optimized by
a low-cost multi-reference calculation that can capture important
contributions from strong electron correlation in their definition.
With regard to (2), the present work does have important connections
to the early refinement of the natural orbitals through iterative
configuration interaction~\cite{D72,TRPB77} and the recent
truncation of natural orbitals in both configuration interaction
calculations~\cite{AS03,BR09} and second-order complete-active-space
perturbation theory~\cite{ATG09}.

While the approach is quite general, here we study the generation of
molecule-optimized basis sets from the solution of the
ACSE~\cite{AC06,*AC07B,*AC07C,*AC07D,ACMR,*GM09b,*RFM09}. By
evolving the ACSE in a large standard atom-centered basis set for a
short distance in the imaginary time-like parameter $\lambda$, we
can generate an approximate 1-RDM whose eigenfunctions provide
approximate natural orbitals. Selection of a smaller set of natural
orbitals based on the occupation numbers generates a
molecule-optimized basis set.  We can choose the rank of this new
basis set equal to that of a smaller standard atom-optimized basis
set which can then be employed to solve the ACSE until convergence.
In this fashion we can generate systematic sets of
molecule-optimized basis sets that significantly accelerate the
solution of multi-reference methods like the ACSE. These basis sets
incorporate important features of chemical bonding and correlation
of the molecule that are not present in the standard atom-optimized
basis sets. Importantly, these molecule-optimized orbitals can be
employed to accelerate any multi-reference quantum chemistry method.

Illustrative applications of the ACSE molecule-optimized basis sets
to the potential energy curves of hydrogen fluoride and diatomic
nitrogen show significant improvements in the nonparallelity errors.
For diatomic nitrogen a molecule-optimized double-zeta-like basis
set yields a nonparallelity error of 0.024~a.u., relative to the TZ
basis set, which significantly improves upon the 0.115~a.u. error in
the DZ basis set.  For hydrogen fluoride the nonparallelity errors
from the ACSE's approximate natural orbitals ordered by occupation
numbers are much better than those from the MCSCF's canonical
orbitals ordered by orbital energies.  Significantly, the
correlation of the active space with the inactive space, as
performed with the approximate solution of the ACSE, is critical to
generating a suitable set of natural orbitals.

While the improvement in absolute energies is not as substantial as
the improvement in the nonparallelity errors, the present results
provide a foundation for future work that may further improve these
results. In section~\ref{sec:ham} we show that basis-set truncation
by approximate natural orbitals can be viewed as a one-electron
unitary transformation of the Hamiltonian operator and suggest an
extension of approximate natural-orbital truncations through {\em
two-electron} unitary transformations of the Hamiltonian operator,
similar to those employed in the ACSE method. In future work we plan
to study larger molecules in larger basis sets, a variety of
approaches for computing approximate sets of natural orbitals,
extrapolations of molecule-optimized basis sets to the
complete-basis-set limit, and extensions of natural-orbital
truncations through two-electron unitary transformations of the
Hamiltonian operator.

The acceleration of the ACSE method for multi-reference correlation
can be applied to extending recent applications of the ACSE to the
study of ground- and excited-state chemical reactions~\cite{FRM09,
GM10} including conical intersections~\cite{SM10,G11,SM11,SM12} in
vinyl alcohol, {\em gauche}-1,3-butadiene, and firefly luciferin.
Often improvements in nonparallelity errors rather than absolute
errors are more important for the accurate prediction of reaction
and excitation energies and other energy differences studied in the
above examples.  The present work can also be applied to correlation
methods that use natural orbitals as their basic variables such as
natural-orbital functional theory~\cite{P07,PML10,P13,SDL08,SDS13,
VGG13}, geminal functional theory~\cite{GFT00, GFT01}, the
precursors of the projected quasi-variational theory~\cite{SJH11},
and the natural-orbital solution of the contracted Schr{\"o}dinger
equation~\cite{M02bb}. The exploitation of molecule-optimized
orbitals and Hamiltonians in electronic structure calculations has
the potential for decreasing computational cost while maintaining
computational accuracy.

\begin{acknowledgments}

DAM gratefully acknowledges the National Science Foundation under
Grant No. CHE-1152425, the Army Research Office under Grant No. W91
INF-1 1-504 1-0085, and Microsoft Corporation for the generous
financial support.  G.G. is supported by an award from the Research
Corporation for Science Advancement and a grant to Gonzaga
University from the Howard Hughes Medical Institute through the
Undergraduate Science Education Program.

\end{acknowledgments}

\bibliography{ACSE_NO_R6}

\newpage

\begin{figure}[htp!]

\includegraphics[scale=0.5]{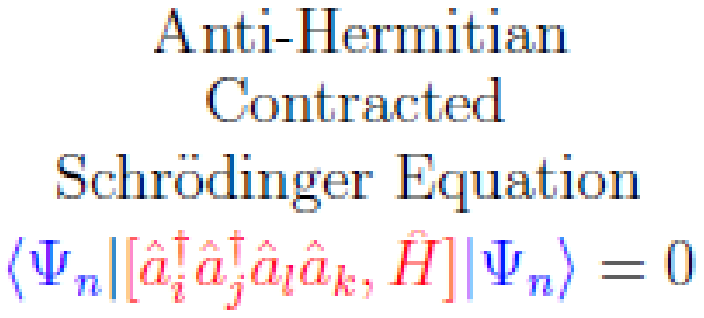}

\caption{Table of Contents Figure}

\end{figure}

\end{document}